\documentclass{article}
\usepackage{graphicx}
\graphicspath{ {./} }

\usepackage[final,nonatbib]{neurips_2019}
\usepackage[utf8]{inputenc} % allow utf-8 input
\usepackage[T1]{fontenc}    % use 8-bit T1 fonts
\usepackage{hyperref}       % hyperlinks
\usepackage{url}            % simple URL typesetting
\usepackage{booktabs}       % professional-quality tables
\usepackage{amsfonts}       % blackboard math symbols
\usepackage{nicefrac}       % compact symbols for 1/2, etc.
\usepackage{microtype}      % microtypography

\title{Deep Discriminative Fine-Tuning for Cancer Type Classification}
\author{
Alena ~Harley\\
Human Longevity Inc.\\
Mountain View, CA 94305\\
\texttt{alenochka@gmail.com} \\
}
\begin{document}

\maketitle

\begin{abstract}
Determining the primary site of origin for metastatic tumors is one of the open problems in cancer care because the efficacy of treatment often depends on the cancer tissue of origin. Classification methods that can leverage tumor genomic data and predict the site of origin are therefore of great value.

Because tumor DNA point mutation data is very sparse, only limited accuracy (64.5\% for 12 tumor classes) was previously demonstrated by methods that rely on point mutations as features~\cite{DeepGene}. Tumor classification accuracy can be greatly improved (to over 90\% for 33 classes) by relying on gene expression data~\cite{RNAClass}. However, this additional data is often not readily available in clinical setting, because point mutations are better profiled and targeted by clinical mutational profiling.

Here we sought to develop an accurate deep transfer learning and fine-tuning method for tumor sub-type classification, where predicted class is indicative of the primary site of origin. Our method significantly outperforms the state-of-the-art for tumor classification using DNA point mutations, reducing the error by more than 30\% at the same time discriminating over many more classes on The Cancer Genome Atlas (TCGA) dataset.  Using our method, we achieve state-of-the-art tumor type classification accuracy of 78.3\% for 29 tumor classes relying on DNA point mutations in the tumor only.
\end{abstract}

\section{Introduction}
Approximately 15\% of cancers metastasize, {\it i.e.} cancer cells break away from where they are first formed (the primary site or tissue of origin) and travel through the blood or lymph system to form new metastatic tumor. Metastatic tumors require further testing to determine the primary site, since the efficacy of cancer treatment is often dependant on the primary site of origin. Some metastatic tumors (4\%) are never fully diagnosed and remain cancer of unknown primary origin. Patients with cancer of unknown primary origin typically have poor survival. Hence, accurate methods that infer the tissue of origin are of great interest.

These methods are also important in the context of blood or urine screening (\textit{i.e.} liquid biopsy) for early detection of cancer. The detection and sequencing of cell-free circulating tumor DNA, as well as circulating tumor cells, has recently been successfully implemented in clinical setting for several cancer types. Once tumor mutations are found in these fluids, methods that can immediately determine the location of the tumor site enable quicker diagnosis and treatment.

Cancer classification using point mutations in tumors is challenging, mainly because the data is very sparse. Many tumours have only a handful of mutations in coding regions, and many mutations are unique, resulting in a long tail of 'private mutations'. It has been previously demonstrated that classifiers that rely on the point mutations in a tumor achieve limited accuracy, particularly 64.5\% on 12 tumor classes~\cite{DeepGene}. More accurate methods for cancer sub-type classification have been developed but they rely on gene expression data that is often not readily available. The accuracy achieved in this setting is over 90\% on 33 tumor classes~\cite{RNAClass}.  Accurate computational methods that can predict tumor class from DNA point mutations alone without relying on additional gene expression data which is not readily available are of great interest. 

Here, we present state-of-the-art deep transfer learning and fine-tuning classification method for tumor sub-type indicative of the primary site of origin. Our method does not require gene expression data and relies on availability of DNA point mutations only.

\section{Methodology}

We used The Cancer Genome Atlas (TCGA) cancer genomic dataset~\cite{TCGA} both for training and testing. Details for the following steps are provided below: 1) the data set and its pre-processing, 2) creation of the gene embedding matrix and encoding tumor samples as images, 3) transfer learning and fine-tuning protocol used for training.

\subsection{Dataset and its pre-processing}

The Cancer Genome Atlas (TCGA) cancer genomic dataset includes 9,642 tumor-normal exome pairs across 33 different cancer sub-types~\cite{TCGA}. We downloaded Mutation Annotation Format (MAF) files from the Genomic Data Commons website (accessed May, 2018)~\cite{GDC}. The colon and rectal cancer cohort (COADREAD), glioblastoma multiforme and lower grade glioma  (GBMLGG) cohort, as well as stomach and esophageal carcinoma cohort (STES) were each treated as single cohort instead of splitting them into sub-cohorts since these are often analyzed together in TCGA studies, thus resulting in 29 cancer sub-types. We removed silent mutations, resulting in a total of 1.3 million non-silent mutations spread across 18,222 genes. The dataset was split -- 80\% of samples within each of the 29 tumor types were used for training and 20\% were used for testing. 

Using training set only we ran MutSigCV~\cite{MutSigCV} to identify significantly mutated genes among the non-silent mutations that were detected in each training set for each tumor type. This let us extract important features of the very sparse dataset. MutSigCV detects genes with higher mutation occurrences than what is expected by chance, taking into account the covariates that include a given gene's base composition, its length, and the background mutation rate. We were left with 1,348 unique significantly mutated genes by setting cut-off to false discovery rate $q<$0.02. 

To learn biologically relevant embedding of the data, we trained Gene2Vec embedding. We utilized database of all known pathways -- MSigDB~\cite{MSigDB} version 6.2, containing 17,810 pathways. In the spirit of Word2Vec~\cite{Mikolov2013}, we mapped pathway-similar genes to nearby points. Here we assumed that genes that appear in the same pathway contexts share biological function. In our implementation we used standard Skip-Gram model when defining Gene2Vec. Gene pairs (33 million) were constructed from the pathway data, and we tried to predict each context gene from its target gene. We used Glorot weight initialization~\cite{Glorot}, and optimized the Noise Contrastive Estimation (NCE) loss function~\cite{Guntmann} (see equation~\ref{eq1}) using Adam~\cite{Kingma2014}.
\begin{equation}
    J_N(\Theta)=\frac{1}{2*N}\sum_n{ln[h(x_n;\Theta)] + ln[1-h(y_n;\Theta)]}
    \label{eq1}
\end{equation}
 Here, $\Theta$ is the set of parameters we optimized, $X= (x_1,...,x_N)$ is the observed data, $Y=(y_1,...,y_N)$ is an artificially generated set of noise, and $h$ denotes the logistic function. We set batch size to 256, number of negative samples to 128, embedding size to 1,348 to match the number of significantly mutated genes, since the goal later is to produce square embeddings of tumor samples. 

\subsection{Transformation of mutation data into images}
\begin{figure}
  \centering
  \includegraphics[scale=0.25]{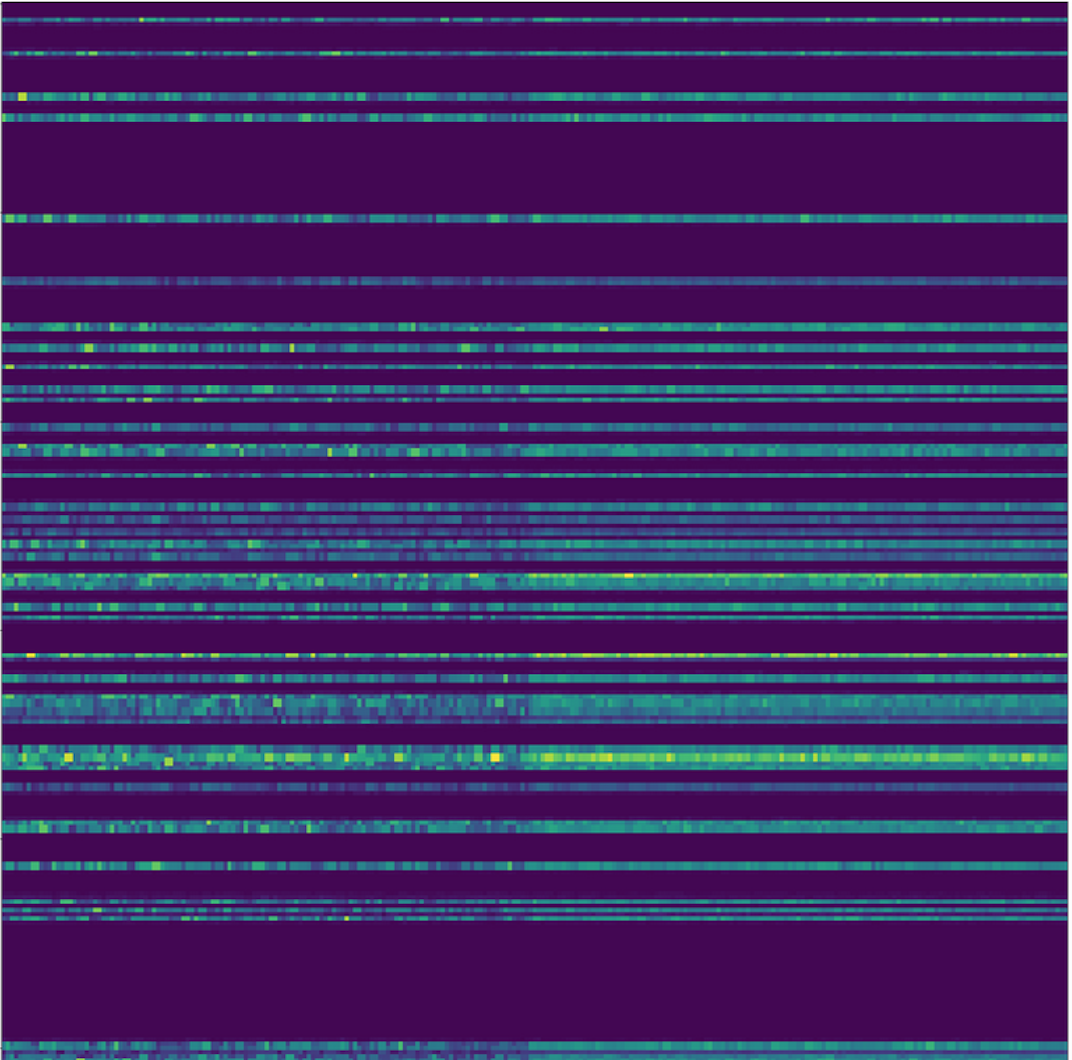}
  %\fbox{\rule[-.5cm]{0cm}{4cm} \rule[-.5cm]{4cm}{0cm}}
  \caption{Example of an embedding for stomach cancer sample.}
  \label{fig1}
\end{figure}

We then extracted learned Gene2Vec embedding for 1,348 significantly mutated genes in our training set, producing a square matrix. TCGA data set is relatively small, so in order to use deep transfer learning methods (trained on images) we used a spectral clustering algorithm~\cite{Stella2003} to create visual structure in the embedding matrix. Spectral clustering is a technique for putting $N$ data points in an $I$-dimensional space into several clusters. Training and test samples were then encoded using spectrally clustered gene embedding using non-silent mutations in the set of significantly mutated genes. Other mutation types were ignored. If a sample did not contain mutations in any of the $1,348$ significantly mutated genes, we queried the embedding to return $10$ closest genes and if any of them were mutated in the sample, the embedding for the closest gene was copied in that row of the matrix. This was done to make our encoding more versatile and to address samples that contain no or very few non-silent mutations in the set of $1,348$ selected genes. The encoded image for significantly mutated genes was replicated in red and blue channels of the final image. The green channel was used to encode both significantly mutated genes and the closest gene embedding if a significantly mutated gene was not altered. An example of an embedding for stomach cancer sample is given in Figure~\ref{fig1}.

\subsection{Transfer learning and fine-tuning training protocol}
Unfortunately, in cancer genomic application domains, training data is scarce, and approaches such as data augmentation are not applicable. Thus, we used transfer representation learning, which can remedy the insufficient training data issue. We pre-train ResNet 34~\cite{ResNet} on ImageNet data set~\cite{ImageNet}. And then used the pre-trained weights as an initialization for the target task of tumor classification utilizing our tumor image embeddings. Images were re-scaled to 512x512 and normalized to match mean and standard deviations of ImageNet images, batch size was set to 32. 

During the first stage of fine tuning all but last custom fully connected layer of the ResNet 34 were frozen~\cite{ULMfit}. The learning rate was chosen to be 0.01 using learning rate finder, see~\cite{CyclicLR} and its implementation in~\cite{FastAI}. The slanted triangular learning rates training schedule~\cite{ULMfit} was used for 10 cycles, and both training and validation loss were still decreasing~\cite{ULMfit,CyclicLR,SuperConv}. 
In the second stage, discriminative fine-tuning~\cite{ULMfit} was used with a sequence of $10^{-6}$ to $10^{-3}$ learning rates. Discriminative fine-tuning splits layers of the deep neural network into groups and applies a different learning rate for each group since different layers should be fine-tuned to different extents; the earliest residual blocks have the smallest, and the fully connected layer has the largest learning rate~\cite{ULMfit}. The learning rate for the last layer $\eta^{L}$ used in stage two was also determined using learning rate finder~\cite{CyclicLR,FastAI}. We  empirically  found  that learning rate $\eta^1=\eta^L/1000$ for the first layer worked best. In stage two of the training  we used slanted triangular learning rates training schedule~\cite{ULMfit} for 12 cycles. 

\section{Results}
\subsection{Quality of gene embedding}
To explore our Gene2Vec embedding, we tested to see if the embedding captured the functional relatedness of genes in terms of their pathway membership. We queried the nearest embedding of a few key cancer genes to spot check the embedding. Since Word2Vec embedding is designed to work well with linear relationships, we examined nearest gene neighbours to NRAS kinase were (cosine similarity of embeddings is listed in parenthesis): HRAS (0.761), KRAS (0.732), PIK3R1 (0.723), MAPK1 (0.722), GRB2 (0.721), AKT1 (0.706), RAF1 (0.704), and MAP2K2 (0.702). The majority of these genes recapitulate the extracellular signal-regulated MAP kinase pathway (RAF/MEK/ERK) that transmits signals from activated cell surface receptors to many cytoplasmic and nuclear targets. The nearest genes for tumor suppressor APC were: CTNNB1 (0.659), PLK1 (0.639), CDKN1A (0.628), PTEN (0.620), CCNB1 (0.620), TP53 (0.620), AKT1 (0.620), and TGFB1 (0.614). These genes recapitulate TGF-Beta signalling pathway which regulates cell proliferation.

\subsection{Tumor classification results}
Table~\ref{tbl1} provides the results of tumor classification. Our deep learning method outperforms the best performing machine learning method. Here, we provide results for XGBoost boosted trees algorithm~\cite{xgb}, since this method was the most competitive. We also ran variants of random forest and support vector classification methods but their performance was worse.

\begin{table}
  \caption{Tumor classification (29 classes) -- experimental results}
  \label{sample-table}
  \centering
  \begin{tabular}{lll}
    \toprule
    \multicolumn{2}{c}{Part}                   \\
    \cmidrule(r){1-2}
    Method     & Accuracy      \\
    \midrule
    XGBoost & $51.2\%$     \\
    ResNet 34 stage 1  & $73.2\%$ \\
    ResNet 34 stage 2  & $78.3\%$  \\
    \bottomrule
  \end{tabular}
  \label{tbl1}
\end{table}

 TCGA tumor cohorts can be generally grouped into the following organ systems: central nervous system (GBMLGG), core gastrointestinal (STES, COADREAD), developmental gastrointestinal (LIHC, PAAD, CHOL), endocrine (THCA and ACC), gynecologic (OV, UCEC, CESC, BRCA), head and neck (HNSC), hematologic and lymphatic malignancies (LAML, DLBC, THYM), melanocytic (SKCM and UVM), neural-crest-derived tissues (PCPG), soft tissue (SARC and UCS), thoracic (LUAD, LUSC), urologic (BLCA, PRAD, TGCT, KIRC, KICH, KIRP)\footnote{See TCGA~\cite{TCGA} for list of abbreviations used.}.

We observed that our mis-classifications are primarily within the same organ systems: ovarian serous cystadenocarcinoma and breast carcinoma; cervical and endocervical cancer and breast carcinoma. We also observed that ovarian serous cystadenocarcinoma was the class with the most errors. This is not surprising since this cancer type have important drivers in the space of non-point mutations -- copy number variants.

\section{Discussion}
Deep neural networks provide the state-of-the-art performance in multiple domains such as images, text, speech. However, in the health and particularly genomic sub-domains there are fewer such examples that outperform other machine learning methods (boosted trees, random forests, and support vector machines). In this paper we describe a way to encode genomic data as images in such a way that transfer learning and fine-tuning can be used to outperform other machine learning methods by a large margin.  

Our three main contributions are: 

\begin{enumerate}
\item Generation of biologically relevant encoding for genomic mutations leveraging pathway information, Gene2Vec embedding, spectral clustering and image creation.
\item An effective training protocol, fine-tuned to the problem at hand, similar training protocol was first introduced in~\cite{ULMfit}. The protocol leverages state-of-the-art transfer learning and fine-tuning techniques.
\item Development of a state-of-the-art classifier for cancer primary site of origin.
\end{enumerate}

As part of the future work, we look forward to 1) improving our understanding of the genes and pathways that are recurrently mutated in cancer by developing better methods to discover significantly mutated genes, 2) integrating DNA copy number data to increase the power to detect new mutational patterns and cancer sub-types, 3) increasing accuracy and addressing more fine-tuned cancer sub-class classification.


\begin{thebibliography}{9}
\bibitem{DeepGene} Yuan, Y., Shi, Y., Li, C., Kim, J., Cai, W., Han, Z., \& Feng, D.D. (2016) DeepGene: an advanced cancer type classifier based on deep learning and somatic point mutations. BMC Bioinformatics, {\bf 17} (Suppl 17):476, \url{https://doi.org/10.1186/s12859-016-1334-9}.
\bibitem{RNAClass} Li, Y., Kang, K., Krahn, J.M., Croutwater, N., Lee, K., Umbach, D.M., \& Li, L. (2017) A comprehensive genomic pan-cancer classification using The Cancer Genome Atlas gene expression data. BMC Genomics, {\bf 18} (1):508, \url{https://doi.org/10.1186/s12864-017-3906-0}.
\bibitem{TCGA} The Cancer Genome Atlas homepage, \url{http://cancergenome.nih.gov/abouttcga}.
\bibitem{GDC} Grossman, Robert L., Heath, Allison P., Ferretti, Vincent, Varmus, Harold E., Lowy, Douglas R., Kibbe, Warren A., Staudt, Louis M. (2016) Toward a Shared Vision for Cancer Genomic Data. New England Journal of Medicine, {\bf 375}:12, 1109-1112, \url{https://gdc.cancer.gov/about-gdc}.
\bibitem{MutSigCV} Lawrence , M. S., Stojanov, P., Polak, P., Kryukov, G. V., et al. (2013) Mutational heterogenieity in cancer and the search for new cancer genes.  Nature, {\bf 499} (7457):214-218, \url{https://doi.org/10.1038/nature12213}.
\bibitem{MSigDB} Subramanian, Tamayo, et al. (2005) Molecular Signatures Database (MSigDB), PNAS, {\bf 102}:15545-15550, \url{http://software.broadinstitute.org/gsea/msigdb/index.jsp}
\bibitem{Mikolov2013} Mikolov, T., Sutskever, I., Chen, K., Corrado, G.S., \& Dean, J. (2013)  Distributed representations of words and phrases and their compositionality. Neural information processing systems.
\bibitem{Glorot} Glorot, X., \& Bengio, Y. (2010) Understanding the difficulty of training deep feedforward neural networks.  In AISTATS, {\bf 9}: 249–256.
\bibitem{Guntmann} Guntman, M., \& Hyvarine,A. (2010) Noise-contrastive estimation: A new estimation principle for unnormalized statistical models. AISTATS, {\bf 9}:297–304.
\bibitem{Kingma2014} Kingma, D.P., \& Ba, J. (2014) Adam: A Method for Stochastic Optimization. ICLR, \url{https://arxiv.org/abs/1412.6980}.
\bibitem{Stella2003}  	Yu, S. X., \&  Shi, J. (2003) Multiclass Spectral Clustering. \url{http://www1.icsi.berkeley.edu/~stellayu/publication/doc/2003kwayICCV.pdf}.
\bibitem{ResNet} He, K., Zhang, X., \& Ren, S. (2016) Deep Residual Learning for Image Recognition. CVPR.
\bibitem{ImageNet} Deng, J., Dong, W., Socher, R., Li, L.-J., Li, K., \& Fei-Fei, L. (2009) ImageNet: A Large-Scale Hierarchical Image Database. CVPR.
\bibitem{ULMfit} Howard, J., \& Ruder, S. (2018) Universal Language Model Fine-tuning for Text Classification. ACL, \url{https://arxiv.org/pdf/1801.06146.pdf}.
\bibitem{CyclicLR} Smith, L.N. (2017) Cyclical Learning Rates for Training Neural Networks. WACV, \url{https://arxiv.org/abs/1506.01186}.
\bibitem{FastAI} FastAi, \url{http://www.fast.ai/}.
\bibitem{SuperConv} Smith, L.N., \& Topin, N. (2017) Super-Convergence: Very Fast Training of Residual Networks Using Large Learning Rates. \url{https://arxiv.org/pdf/1708.07120.pdf}.
\bibitem{tSNE} van der Maaten, L., \& Hinton, G. (2008) Visualizing data using t-SNE. The Journal of Machine Learning Research, {\bf 9}:2579-2605.
\bibitem{xgb} Chen, T., Guestrin, C. (2016) XGBoost: A Scalable Tree Boosting System. SIGKDD, 785-794. 
\end{thebibliography}
\end{document}